\documentclass[conference, a4paper]{IEEEtran}
\IEEEoverridecommandlockouts
\usepackage{multirow}
\usepackage{cite}
\usepackage{inputenc}
\usepackage{amsmath,amssymb,amsfonts}
\usepackage{algorithmic}
\usepackage{graphicx}
\usepackage{textcomp}
\usepackage{color}
\usepackage{setspace}
\def\BibTeX{{\rm B\kern-.05em{\sc i\kern-.025em b}\kern-.08em
		T\kern-.1667em\lower.7ex\hbox{E}\kern-.125emX}}
\usepackage{subfigure}
\newtheorem{my_theorem}{Theorem}

\title{Optical Wireless Transmissions over Multi-layer Underwater Channels with Generalized Gamma Fading}

\author{\IEEEauthorblockN{Suhrid Das}
	\IEEEauthorblockA{Department of Electronics and\\
		Communication Engineering\\
		Jalpaiguri Government Engineering College\\
		Jalpaiguri-735102, West Bengal, India\\
		suhriddas2000@gmail.com}
	\and
	\IEEEauthorblockN{Ziyaur Rahman}
	\IEEEauthorblockA{Department of Electrical and\\
		Electronics Engineering\\
		BITS Pilani, Pilani Campus\\
		Pilani-333031, Rajasthan\\
		p20170416@pilani.bits-pilani.ac.in}
	\and
	\IEEEauthorblockN{S.~M.~Zafaruddin}
	\IEEEauthorblockA{Department of Electrical and\\
		Electronics Engineering\\
		BITS Pilani, Pilani Campus\\
		Pilani-333031, Rajasthan\\
		syed.zafaruddin@pilani.bits-pilani.ac.in}	
		\thanks{This work was supported in part by the Science and Engineering Research Board, Department of Science and Technology (DST), India under Grant MTR/2021/000890, and  SRG/2019/002345.}
}

\begin{document}
	\maketitle
\begin{abstract}
Underwater optical communication (UWOC) is a potential solution for broadband connectivity in oceans and seas for underwater applications  providing high data rate transmission with low latency and high reliability.  Recent measurement campaigns suggest generalized Gamma distribution as a viable model for oceanic turbulence.  In this paper, we analyze the performance of a UWOC system by modeling the vertical underwater link as a multi-layer cascaded channel, each distributed according to independent but not identically distributed (i.ni.d.) generalized Gamma random variables and considering the zero bore-sight model for pointing errors. We derive analytical expressions for probability density function (PDF) and cumulative distribution function (CDF) for the signal-to-noise
ratios (SNR) of the combined channel and  develop  performance metrics of the considered UWOC system using outage probability, average bit error rate (BER), and ergodic capacity.  We also derive the asymptotic expressions for outage probability and average BER to determine the diversity order of the proposed system for a better insight into the system performance. We use Monte-Carlo simulation results to validate our exact and asymptotic expressions and demonstrate the performance of the considered underwater UWOC system  using measurement-based parametric data available for  turbulent oceanic  channels.
\end{abstract}


\section{Introduction}
 Optical wireless communication (OWC) is a promising technology for  underwater data transmission  providing higher throughput with low latency and high reliability than radio frequency (RF) and acoustic wave communication systems. The underwater optical communication (UWOC) system transmits  data in an  unguided water environment using the wireless optical carrier   for military, economic and scientific applications \cite{Zhaoquan2017}. Despite several advantages of the UWOC,  the underwater link suffers from  signal attenuation, oceanic turbulence, and pointing errors.  The signal attenuation occurs due to the  molecular absorption and scattering effect of each photon propagating through water, generally modeled by the extinction coefficient. Oceanic turbulence is the effect of random variations in the refractive index of the  UWOC channel caused by  random variations of water temperature, salinity, and air bubbles \cite{Vahid2018}.  Pointing errors can also be detrimental to UWOC transmissions  due to misalignment between the transmitter and detector apertures. Therefore, it is  desirable to analyze the UWOC systems over various underwater channel impairments  for an effective system design.

There has been tremendous research on the performance assessment of UWOC systems in recent years \cite{Xiaobin2020,Gercekcioglu2014,Vahid2016,Mingjian2016,Mandeep2018,Vahid2015,Jamali2017,Azadeh2017}. The  authors in \cite{Xiaobin2020}  provided an overview of  various challenges associated with UWOC and proposed positioning, acquisition, and tracking scheme to mitigate the effect of pointing errors under turbulent channels. The average bit-error-rate (BER) performance under weak log-normal distributed turbulence channels was  presented in  \cite{Gercekcioglu2014}. The authors in \cite{Vahid2016} characterized a relay-assisted UWOC with optical code division multiple access (OCDMA) over log-normal turbulent channels. An analytic expression for the channel capacity of an orbital angular momentum (OAM) based free-space optical (FSO) communication in weak oceanic turbulence was developed in  \cite{Mingjian2016}. The effect of air bubbles on the UWOC was experimentally evaluated in  \cite{Mandeep2018}.  The authors in \cite{Vahid2015,Vahid2017}  analyzed the performance of  multi-input and multi-output (MIMO) UWOC systems over  log-normal turbulent channels.   Further, a multihop UWOC system was investigated in \cite{Jamali2017}. The outage probability of a multiple decode-and-forward (DF) relay-assisted  UWOC system with an on-off keying (OOK) modulation  was studied in \cite{Azadeh2017}.

In those mentioned above and related research, a single layer of oceanic turbulence channel over the  entire transmission range has been considered. However, experimental results reveal ocean stratification, i.e., the temperature gradient and salinity are depth-dependent (typically varying between a few meters to tens of meters), resulting in  many non-mixing  layers with different oceanic turbulence \cite{Elamassie2019}.  Thus, considering multiple oceanic layers for vertical transmissions may  provide a more realistic performance assessment for UWOC systems. In \cite{Elamassie2018,Mohammed2018,Elamassie2019,Elamassie2019pe},  the author analyzed the performance vertical UWOC links by cascading the end-to-end link as the concatenation of multiple layers considering both log-normal and Gamma-Gamma oceanic turbulent channels for each layer. In \cite{Vahid2018}, the authors presented a holistic experimental view on the statistical characterization of oceanic turbulence in UWOC systems, considering the effect of the temperature gradient, salinity, and air bubbles. They used various statistical distributions such as log-normal, Gamma, Weibull, Exponentiated Weibull, Gamma-Gamma, and generalized Gamma  to model  underwater turbulence channels. Experimental investigation in \cite{Vahid2018} projected  the  generalized Gamma distribution as a more generic  model and was valid for various  underwater channel conditions. To the best of the authors’ knowledge, there are no analyses available for the outage probability, average BER, and ergodic capacity of a multi-layer UWOC system over a generalized Gamma turbulent channel with pointing errors.

In this paper, we analyze the performance of a vertical UWOC system under the combined effect of cascaded underwater turbulence channels and pointing errors. The major contributions of the proposed work are summarized as follows:
\begin{itemize}
\item We use Mellin transform to derive the novel probability density function (PDF) and cumulative distribution function (CDF) of the signal-to-noise ratio (SNR) for vertical UWOC link in terms of a single-variate Fox H-function considering independent and not identically distributed (i.ni.d.) generalized Gamma distribution model for underwater turbulence channel and zero bore-sight  model  for pointing errors.

\item We use the derived statistical results to develop analytical expressions for the outage probability, average BER, and ergodic capacity of the cascaded UWOC system.
\item We also present an asymptotic analysis for both outage probability and average BER in the high SNR regime to derive the diversity order depicting the impact of system and channel parameters on the performance of the considered system. 
\item We use numerical and simulation analysis to validate our derived expressions and demonstrate the performance of the considered UWOC system for various parameters of interest.
\end{itemize}
\begin{figure}[tp]	
	\centering
	\includegraphics[scale=0.4]{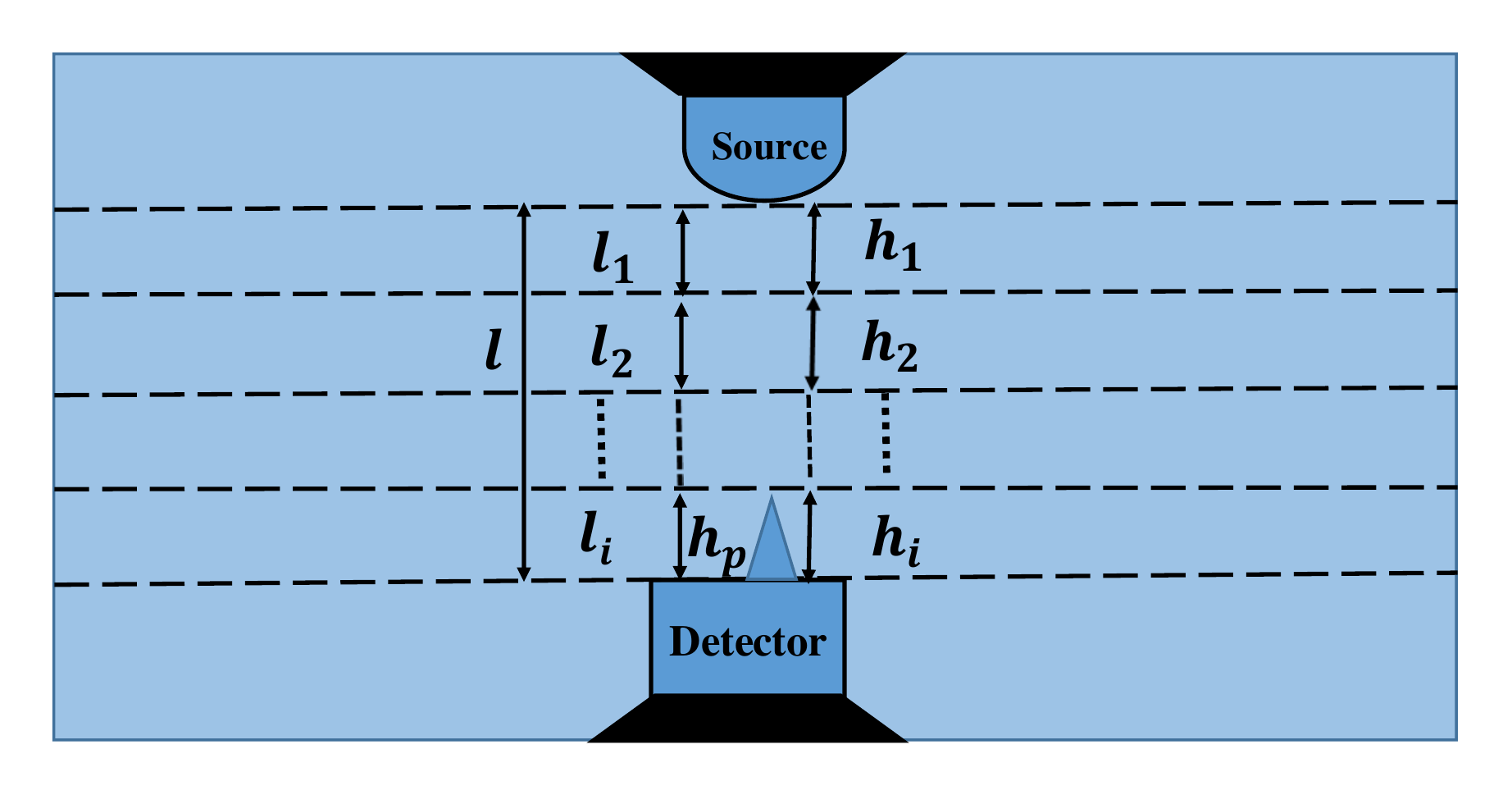}
	\vspace{-0.6cm}
	\caption{ Schematic diagram  for multi-layer UWOC system.}
	\label{system_model}	
\end{figure}

\section{System Model}
We consider a UWOC system by splitting  the entire transmission channel in $N$ distinct layers in succession, resulting in $N$ vertical links,  as shown in Fig~\ref{system_model}. We use the non-coherent intensity modulation/direct detection (IM/DD) scheme, where the photodetector detects changes in the light intensity without employing a local oscillator. The  heterodyne detection (HD) requires complex processing of  mixing  the received signal with a coherent signal produced by the local oscillator \cite{Melchior1970}. Thus, the received electrical signal for  the transmitted signal   $s$ under   additive white gaussian noise (AWGN) $w$  with variance $\sigma_{w}^2$ can be expressed as
\begin{eqnarray}
    y=h_lh_ch_{p}s+ w
    \label{eq: System Model main eq}
\end{eqnarray}
where $h_l=e^{-\alpha l}$ is the atmospheric path gain with link distance $l$ (in \rm{m}) and extinction attenuation coefficient $\alpha$,  the term $h_c=\prod_{i=1}^{N}h_{i}$ (where $i=1,2,3\cdots$ $N$) is the cascaded channel with $h_i$ as the $i$-th layer of vertical link, and   $h_{p}$ models  pointing errors. The fading coefficients $h_i$, $i=1,2, \cdots, N$ associated with the different layers are modeled using i.ni.d generalized Gamma random variables  \cite{Vahid2018}:

\begin{flalign}
	f_{h_{i}}(h_{i})=\frac{p_{i}}{{a_{i}}^{d_{i}}\Gamma\Big(\frac{d_{i}}{p_{i}}\Big)}{h_{i}}^{d_{i}-1}\exp\bigg(-\Big(\frac{h_{i}}{a_{i}}\Big)^{p_{i}}\bigg), 0 \leq h_{i} \leq \infty
	\label{eq:h_gg}
\end{flalign}
where $\Gamma(z)=\int_{0}^{\infty}x^{z-1}\exp(-x)dx$ denotes the Gamma function. Here, $a_{i}$, $d_{i}$, and $p_{i}$ are distribution parameters for the $i$-th layer to model different oceanic turbulence scenarios, as given in \cite{Vahid2018} (see Table-I, Table-II, and Table-III). As such, $p_i=1$ in 	\eqref{eq:h_gg} denotes a Gamma distribution representing  a thermally uniform UWOC channel. 

Assuming IM/DD technique and on-off keying (OOK) modulation with $x\in \{0,\sqrt{2}P_t\}$  and $P_t$ as average transmitted optical, the instantaneous  received electrical SNR is given by \cite{Farid2007}
\begin{eqnarray}
\gamma=\frac{P^2_t h^2_l h^2}{\sigma_{w}^2}=\gamma_0 h^2
\label{eq: inst_snr}
\end{eqnarray} 
where $h=h_ch_p$ is the combined channel and $\gamma_0=\frac{P^2_t h^2_l}{\sigma_{w}^2}$ is the average electrical SNR. Note that  $P_t^2$ in \eqref{eq: inst_snr} is attributed to the detection type IM/DD and becomes $\gamma=\frac{P_t h_l h}{\sigma_{w}^2}$ for the HD technique \cite{Melchior1970, Chapala2021}.

The PDF of zero-boresight pointing errors fading $h_{p}$ is given in \cite{Farid2007}:
\begin{eqnarray}
f_{h_{p}}(h_p) = \frac{\rho^2}{A_{0}^{\rho^2}}h_{p}^{\rho^{2}-1},0 \leq h_{p} \leq A_0
\label{eq:pdf_hp}
\end{eqnarray}
where $A_0=\mbox{erf}(\upsilon)^2$ with $\upsilon=\sqrt{\pi/2}\ r/\omega_{z}$, $r$ is the aperture radius and $\omega_{z}$ is the beam width, and $\rho = {\frac{\omega_{{z}_{\rm eq}}}{2 \sigma_{s}}}$ with  $\omega_{{z}_{\rm eq}}$ as the equivalent beam width at the receiver and $\sigma^2_{s}$ as the variance of pointing errors displacement characterized by the horizontal sway and elevation.

\section{Performance Analysis}
In this section, we analyze the performance of a cascaded UWOC system by deriving  exact expressions for the PDF and CDF of the SNR under the combined effect of oceanic turbulence and pointing errors and provide expressions for the outage probability, average BER, and ergodic capacity. The derived performance assessment can help  network operators design efficient UWOC systems when the underwater transmission range is high.

\subsection{PDF and CDF of SNR}
Similar to \cite{Vahid2018}, we assume that turbulence channels are independent for each layer to provide a tractable performance analysis. However,   adjacent layers might become correlated if the layers are not sufficient apart, requiring rigorous analysis for the product of $N$ correlated random variables.

Denoting $z_1=x_1x_2$  as the product of two i.i.d. generalized Gamma distributions, we use \cite[pp. $368$, eq. $3.471.09$]{Zwillinger2014} to get the PDF for two-layers UWOC system:
	\begin{equation}
		f_{z_1}(z)=\frac{2}{p}K_0\left(2\sqrt{\frac{z^p}{a^{2p}}} \right)
		\label{eq:PDF_2_channel1}
	\end{equation}
	where $K_0$ denotes the Bessel function of the zeroth order. It can be seen that \eqref {eq:PDF_2_channel1} does not represent a pattern comparing with \eqref{eq:h_gg}, and thus cannot be generalized to derive the PDF for the product of $N$ channels. In the following theorem, we use Mellin's inverse transform to provide PDF and CDF  of  the cascaded UWOC link under the combined effect of the oceanic turbulence and pointing errors under i.ni.d fading channels. 
\begin{my_theorem}
The PDF and CDF of the end-to-end SNR for $N$ cascaded channels distributed according to  generalized Gamma channel combined with pointing errors are given as
\begin{eqnarray}
    &f_\gamma(\gamma)=\frac{\rho^2}{2\gamma}\prod_{i=1}^{N} \frac{1}{\Gamma\left(\frac{d_i}{p_i} \right)} H_{1,N+1}^{N+1,0} \nonumber \\&\left[\begin{matrix} (1+\rho^2,1) \\\left\{\left(\frac{d_{i_1}}{p_{i_1}},\frac{1}{p_{i_1}} \right)\right\}_{i_1=1}^{N}, (\rho^2,1) \end{matrix} \bigg|\prod_{{i_2}=1}^{N} \left(\frac{1}{a_{i_2}A_0}\sqrt{\frac{\gamma}{\gamma_0}}\right)\right]
     \label{eq:pdf_hc_hp}
\end{eqnarray}

\begin{flalign}
    &F_{\gamma}(\gamma)=\prod_{i=1}^{N} \frac{\rho^2}{\Gamma\left(\frac{d_i}{p_i} \right)} H_{2,N+2}^{N+1,1} \nonumber \\&\left[\begin{matrix} (1,1), (1+\rho^2,1)  \\\left\{\left(\frac{d_{i_1}}{p_{i_1}},\frac{1}{p_{i_1}} \right)\right\}_{{i_1}=1}^{N}, (\rho^2,1), (0,1)\end{matrix} \bigg|\prod_{{i_2}=1}^{N} \left(\frac{1}{a_{i_2}A_0}\sqrt{\frac{\gamma}{\gamma_0}}\right)\right]
     \label{eq:cdf_hc_hp}
\end{flalign}
\end{my_theorem}

\begin{IEEEproof}
See Appendix A.
\end{IEEEproof}

In what follows, we use Theorem 1 to develop performance metrics of the UWOC system.
\subsection{Outage Probability}
Outage probability is a performance metric that demonstrates the effect of fading channels on the communication systems.  It is defined as the probability that the instantaneous  SNR falls below a certain threshold $\gamma_{th}$ and is given as
\begin{eqnarray}
   	P_{\text{out}}(\gamma_{\text{th}},\gamma_{0})=P(\gamma<\gamma_{\text{th}})=F_{\gamma}(\gamma_{\text{th}})
    \label{eq:outage_prob_general}
\end{eqnarray}
Substituting \eqref{eq:cdf_hc_hp} in \eqref{eq:outage_prob_general} yields
\begin{flalign}
  &P_{\rm{out}}=\prod_{i=1}^{N} \frac{\rho^2}{\Gamma\left(\frac{d_i}{p_i} \right)} H_{2,N+2}^{N+1,1}\nonumber \\&\left[\begin{matrix} (1,1), (1+\rho^2,1)  \\\left\{\left(\frac{d_{i_1}}{p_{i_1}},\frac{1}{p_{i_1}} \right)\right\}_{i_1=1}^{N}, (\rho^2,1), (0,1)\end{matrix} \bigg|\prod_{i_2=1}^{N} \left(\frac{1}{a_{i_2}A_0}\sqrt{\frac{\gamma_{th}}{\gamma_0}}\right)\right]
 \label{eq:outage_prob_combine}
\end{flalign}
We  use  \cite[eq. $1.8.4$]{Kilbas} to develop an asymptotic expression for the  outage probability in the high SNR regime $\gamma_0\to \infty$:
\begin{eqnarray}
	&P_{\rm{out}}^{\infty}=\prod_{i=1}^{N} \frac{\rho^2}{\Gamma\left(\frac{d_i}{p_i} \right)}\sum_{k=1}^{N+1}\nonumber\\&\frac{\left(\frac{1}{a_iA_0}\sqrt{\frac{\gamma_{th}}{\gamma_0}}\right)^{\frac{b_k}{\beta_k}}\prod_{j=1, j\neq k}^{N+1}\Gamma\left(b_j-b_k\frac{\beta_j}{\beta_k}\right)\Gamma\left(\frac{b_k}{\beta_k}\right)}{\beta_k\Gamma\left(1+\rho^2-\frac{b_k}{\beta_k}\right)\Gamma\left(1+\frac{b_k}{\beta_k}\right)}
\label{eq:outage_prob_asymp}
\end{eqnarray}
where $b_j=b_k=\{\frac{d_i}{p_i}, \rho^2\}$ and $\beta_j=\beta_k=\{\frac{1}{p_i}, 1\}$. Combining  the exponent of $\gamma_0$ in  \eqref{eq:outage_prob_asymp}, the diversity order of the considered system can be expressed as ${DO}_{{\rm out}}=\sum_{i=1}^{N}\min\{\frac{d_i}{2}, \frac{\rho^2}{2}\}$. 

The diversity order reveals that the outage probability at a high SNR is dependent only on the parameter $d$ of the oceanic turbulence  channel and pointing errors. A sufficient higher beam-width can make the diversity order independent of pointing errors.
\begin{table*}[t]	
	\renewcommand{\arraystretch}{1.1}
	\caption{Simulation Parameters}
	\label{table:simulation_parameters}
	\centering
	\begin{tabular}{|c|c|c|}
		\hline 	
		Transmitted optical power &$P_t$ & $-10$ to $55$ \mbox{dBm} \\ \hline
		
		AWGN variance &$\sigma_w^2$ & $10^{-14}~\rm {A^2/GHz}$ \\ \hline	
		Total link distance &$l$ & $50$ \mbox{m}\\ \hline
		Extinction coefficient& $\alpha$ & $0.056$ \cite{Vahid2018} \\ \hline
		Pointing errors parameters& $A_0$, $\rho^2$ & $0.0032$, $\{1, 6\}$
		\\ \hline
		& $\{a_i\}_{i=1}^5$ &  $\{0.6302, 1.0750, 1.0173, 0.7598, 1.0990\}$\\Generalized Gamma parameters \cite{Vahid2018}& $\{d_i\}_{i=1}^5$ &  $\{1.1780, 3.2048, 1.6668, 2.3270, 4.5550\}$ \\ & $\{p_i\}_{i=1}^5$ &  $\{0.8444, 2.9222, 1.0380, 1.4353, 4.6208\}$\\ \hline
		Modulation parameters& $M$, $\delta$, $\phi$, $q_n$ & $1$, $1$, $\frac{1}{2}$, $\frac{1}{2}$
		\\ \hline	
	\end{tabular}
	\label{Simulation_Parameters}	 
\end{table*} 

\subsection{Average BER}
In this subsection, we analyze  the average BER performance of the cascaded UWOC system. The average BER can be obtained as \cite{dual_hop_turb2017}:
\begin{eqnarray} 
\bar{P}_{e} = \frac{\delta}{2\Gamma(\phi)}\sum_{n=1}^{M}q_n^{\phi}\int_{0}^{\infty} \gamma^{\phi-1} {\exp(-q_n \gamma)} F_{\gamma} (\gamma)   d\gamma
\label{eq:ber}
\end{eqnarray}
where the set $\{M, \delta, \phi, q_n\}$ can specify a variety of modulation schemes.
Using \eqref{eq:cdf_hc_hp} and substituting    $\exp(-q_n \gamma)=G_{0,1}^{1,0} [\begin{matrix} - \\ 0\end{matrix} \bigg| q_n \gamma]$ in \eqref{eq:ber} and representing the Meijer G-function into Fox-H function, we express   \eqref{eq:ber} as
\begin{eqnarray}
&\bar{P_e}=\frac{\delta\rho^2}{2\Gamma(\phi)}\sum_{n=1}^{M}q_n^{\phi}\prod_{i=1}^{N} \frac{1}{\Gamma\left(\frac{d_i}{p_i}\right)}\nonumber \\& \times \int_{0}^{\infty}\gamma^{\phi-1} H_{0,1}^{1,0} \left[\begin{matrix}  - \\ (0,1)\end{matrix} \bigg| q_n \gamma \right] H_{2,N+2}^{N+1,1} \nonumber \\&\left[\begin{matrix}(1,1), (1+\rho^2,1) \\\left\{\left(\frac{d_{i_1}}{p_{i_1}},\frac{1}{p_{i_1}} \right)\right\}_{i_1=1}^{N}, (\rho^2,1), (0,1)\end{matrix} \bigg|\prod_{i_2=1}^{N} \left(\frac{1}{a_{i_2}A_0}\sqrt{\frac{\gamma}{\gamma_0}}\right)\right]d\gamma \nonumber \\&
\label{eq:avg_BER_combined}
\end{eqnarray}
We apply the definite integral of product of two Fox-H functions in \cite[eq. $2.8.4$]{Kilbas} to express the average BER as
\begin{eqnarray}
&\bar{P_e}=\frac{\delta\rho^2}{2\Gamma(\phi)}\sum_{n=1}^{M}\prod_{i=1}^N\frac{1}{\Gamma \left(\frac{d_i}{p_i}\right)}H_{3,N+2}^{N+1,2}\nonumber \\& \left[\begin{matrix} (1,1), (1-\phi,\frac{1}{2}), (1+\rho^2,1) \\\left\{\left(\frac{d_{i_1}}{p_{i_1}},\frac{1}{p_{i_1}} \right)\right\}_{i_1=1}^{N}, (\rho^2,1), (0,1) \end{matrix} \bigg|\prod_{i_2=1}^{N} \frac{1}{a_{i_2}A_0\sqrt{q_n \gamma_0}}\right]\nonumber \\&
\label{eq:avg_BER_combined_final}
\end{eqnarray}

We apply the series expansion in \cite[eq. $1.8.4$]{Kilbas} to get an asymptotic expression for the average BER at a high SNR $\gamma_0\to \infty$:
\begin{eqnarray}
&\bar{P_e}^{\infty}=\frac{\delta\rho^2}{2\Gamma(p)}\sum_{n=1}^{M}\prod_{i=1}^N\frac{1}{\Gamma \left(\frac{d_i}{p_i}\right)}\sum_{k=1}^{N+1}\nonumber\\&\frac{\left(\frac{1}{a_iA_0\sqrt{q_n \gamma_0}}\right)^{\frac{b_k}{\beta_k}}\prod_{j=1,j\neq k}^{N+1}\Gamma\left(b_j-b_k\frac{\beta_j}{\beta_k}\right)\Gamma\left(\frac{b_k}{\beta_k}\right)\Gamma\left(\phi+\frac{b_k}{2\beta_k}\right)}{\beta_k\Gamma\left(1+\rho^2-\frac{b_k}{\beta_k}\right)\Gamma\left(1+\frac{b_k}{\beta_k}\right)}
\label{eq:avg_BER_asymp}
\end{eqnarray}
where $b_j=b_k=\{\frac{d_i}{p_i}, \rho^2\}$ and $\beta_j=\beta_k=\{\frac{1}{p_i}, 1\}$.
Similar to the outage probability, the diversity order for the average BER  can be obtained as  ${DO}_{{\bar{P}_e}}=\sum_{i=1}^{N}\min\{\frac{d_i}{2}, \frac{\rho^2}{2}\}$. Thus turbulent channl parameter $d_i$ and pointing error parameter $\rho^2$ determines the diversity order of the system. Thus, the effect of pointing errors can be mitigated using a sufficient higher beam-width.
\subsection{Ergodic Capacity}
The ergodic capacity $\bar{C}$ is an important performance metric for  the design of communication systems, and it can be defined as \cite{Nistazakis2009}:
\begin{eqnarray}
\bar{C}&=\int\limits_{0}^{\infty} \log_2(1+\kappa\gamma) f_\gamma(\gamma) d\gamma
\label{eq:general_capacity_exp}
\end{eqnarray}
where $\kappa=\frac{e}{2 \pi}$ for IM/DD  and $\kappa=1$ for the HD technique. Using \eqref{eq:pdf_hc_hp} and substituting $ \log_2(1+\kappa\gamma)=1.44 G_{2,2}^{1,2} \left[\begin{matrix} 1,1 \\ 1,0\end{matrix} \bigg| \kappa\gamma \right]$ in \eqref{eq:general_capacity_exp} and representing  Meijer-G function into Fox-H function, we get
\begin{flalign}
     &\bar{C}=0.72\rho^2\prod_{i=1}^{N} \frac{1}{\Gamma\left(\frac{d_i}{p_i}\right)} \int_{0}^{\infty}\gamma^{-1} H_{2,2}^{1,2} \left[\begin{matrix} (1,1), (1,1) \\ (1,1), (0,1) \end{matrix} \bigg| \kappa\gamma \right]\nonumber \\& H_{1,N+1}^{N+1,0} \left[\begin{matrix} (1+\rho^2,1) \\\left\{\left(\frac{d_{i_1}}{p_{i_1}},\frac{1}{p_{i_1}} \right)\right\}_{i_1=1}^{N} (\rho^2,1) \end{matrix} \bigg|\prod_{i_2=1}^{N} \left(\frac{1}{a_{i_2}A_0}\sqrt{\frac{\gamma}{\gamma_0}}\right)\right]d\gamma \nonumber \\&
     \label{eq:capacity_combine1}
\end{flalign}
Thus, we apply the definite integral of product of two Fox-H functions in \cite[eq. $2.8.4$]{Kilbas} to get  an analytical expression for the ergodic capacity of  the considered UWOC system:
\begin{flalign}
    &\bar{C}=0.72\rho^2\prod_{i=1}^{N} \frac{1}{\Gamma\left(\frac{d_i}{p_i}\right)}H_{3,N+3}^{N+3,1}\nonumber \\& \left[\begin{matrix} (0,\frac{1}{2}),(1,\frac{1}{2}),(1+\rho^2,1) \\\left\{\left(\frac{d_{i_1}}{p_{i_1}},\frac{1}{p_{i_1}} \right)\right\}_{i_1=1}^{N},(\rho^2,1)(0,\frac{1}{2}),(0,\frac{1}{2}) \end{matrix} \bigg|\prod_{i_2=1}^{N} \frac{(\kappa\gamma_0)^{-\frac{1}{2}}}{a_{i_2}A_0}\right]\nonumber \\&
    \label{eq:capacity_combine_final}
\end{flalign}
\section{Simulation and numerical analysis}
In this section, we demonstrate the performance of vertical UWOC system over-generalized Gamma fading with pointing errors. We use Monte-Carlo (MC) simulation (averaged over $10^7$ channel realizations) to validate the derived analytical expressions. We use standard inbuilt MATLAB and MATHEMATICA libraries to calculate Meijer-G and Fox-H function, respectively. Since there is no measurement data to confirm the variation of distribution parameters with distance, we illustrate the performance by considering vertical underwater link length $l=50$ {\rm{m}} with $N=5$ layers and the thickness of each layer is assumed to be $10$ {\rm{m}}. Other  simulation parameters are listed  in Table \ref{Simulation_Parameters}.

\begin{figure}[t]
	{\includegraphics[width=\columnwidth]{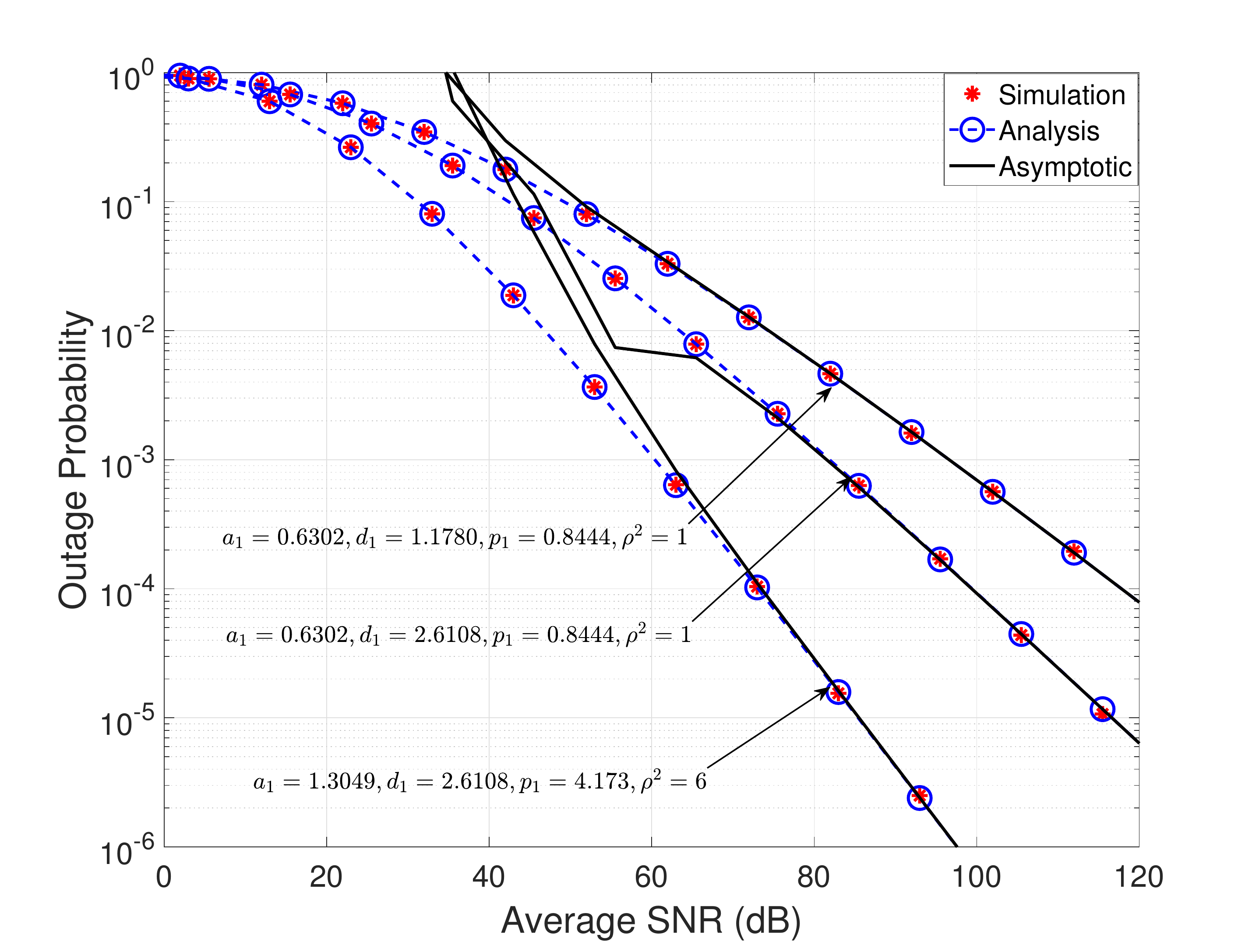}}
	\caption{Outage probability performance for a $5$-layer UWOC.}
	\label{out_prob}
\end{figure}

First, we demonstrate the outage probability performance of the UWOC system in Fig.~\ref{out_prob}. It can be seen from the figure that the outage performance of the system improves with an increase in the values of generalized Gamma distribution parameters ($a$, $d$, and $p$) and a decrease in  pointing errors (i.e., higher $\rho^2$). In the first plot of Fig.~\ref{out_prob}, we consider the pointing errors parameter ($\rho^2=1$) and the generalized Gamma distribution parameters ($a$, $d$, and $p$), as given in Table \ref{Simulation_Parameters}. The diversity order ${DO}_{\rm out}=\sum_{i=1}^{N}\min\{\frac{d_i}{2}, \frac{\rho^2}{2}\}$ for the top and middle plots in  Fig.~\ref{out_prob} are given by $\min\{6.4658, 0.5\}$ and $\min\{7.1822, 0.5\}$, respectively. It can be clearly observed that the diversity order is dependent on the pointing error parameter ($\rho^2$) since the slope does not change  with the oceanic channel parameter $d_i$. Further, in the third plot,  the diversity order becomes $\min\{7.1822, 3\}$, demonstrating a  change of slope with  $\rho^2$, thus confirming our diversity order analysis.

\begin{figure}[t]
	{\includegraphics[width=\columnwidth]{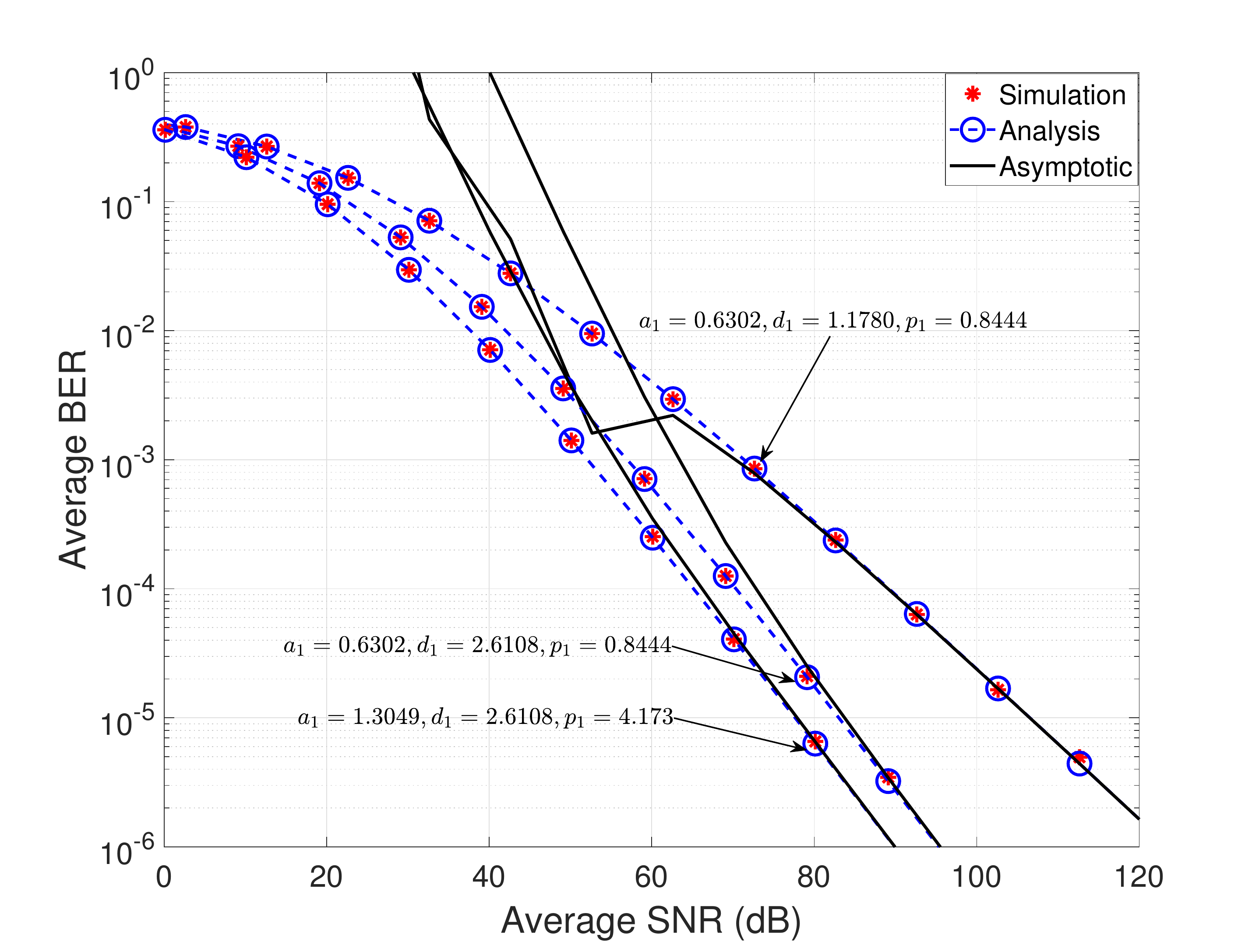}}
	\caption{Average BER performance of UWOC with $\rho^2=6$.}
	\label{avg_ber}
\end{figure}

Next, we present the average BER performance of the vertical UWOC system in Fig.~\ref{avg_ber}. The average BER of the system  follows a similar trend as observed for the outage probability  with turbulent channel and pointing error parameters. It is evident from the plots that the average BER of the system improves by almost ten times if the turbulent  channel parameter $d_i$ increases from $1.1780$ to $2.6108$ at an average SNR of $80$ \mbox{dB}. Further, the diversity order follows a similar analysis as that of the outage probability, which can be confirmed by observing  the  slope change among the plots.


\begin{figure}[t]
	{\includegraphics[width=\columnwidth]{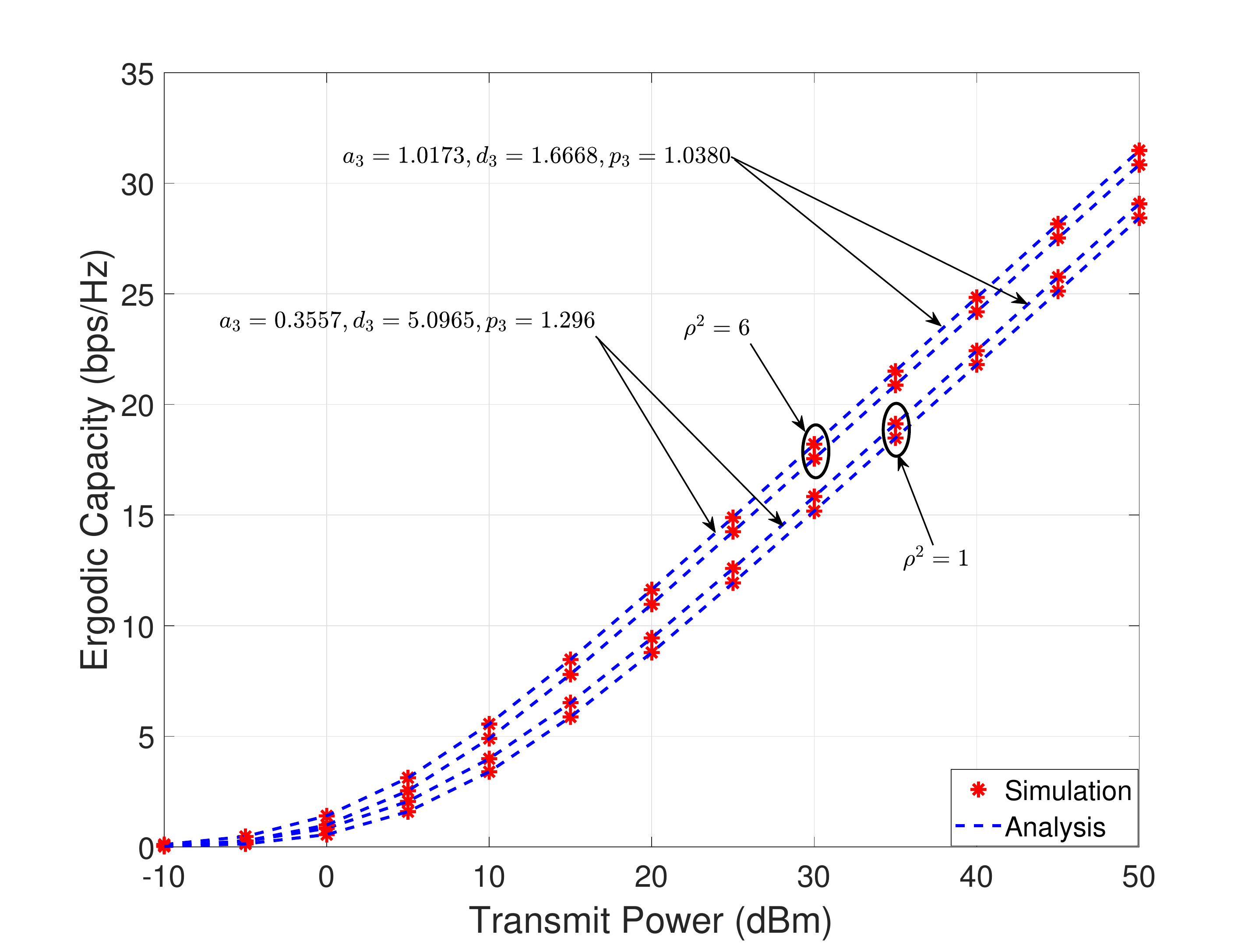}}
	\caption{Ergodic capacity performance of UWOC system.}
	\label{erg_cap}
\end{figure}
Finally, we demonstrate the effects of the different vertical link impairments on the UWOC system by plotting the ergodic capacity versus transmit power, as shown in Fig.~\ref{erg_cap}. The figure shows that the ergodic
capacity increases by almost $3$ \mbox{bits/sec/Hz} if the effect of pointing error decreases by increasing the parameter $\rho^2$ from $1$ to $6$ for the given turbulent  parameters. It can also been seen that the ergodic capacity increases by almost $1$ \mbox{bits/sec/Hz} if we change the oceanic turbulence parameters ($a_3=0.3557$, $d_3=5.0965$, and $p_3=1.296$) for the both  $\rho^2=1$ and $\rho^2=6$.

In all the above plots (Fig.~\ref{out_prob}, Fig.~\ref{avg_ber}, and Fig.~\ref{erg_cap}) we verify our derived results using MC simulation, demonstrating that our derived analytical expressions (depicted as ‘Analysis’) of outage probability, average BER, and ergodic capacity have an excellent match with MC simulations (depicted as ‘Simulation’). Further,  the asymptotic expression of outage probability and average BER can be see to converge with analysis and simulation results in the high SNR regime.

\section{Conclusion} 
In this paper, we analyzed the performance of the UWOC system considering the vertical underwater link as a multi-layer cascaded channel, each distributed according to i.ni.d. generalized Gamma random variables. We analyzed the system performance by deriving analytical expressions of the PDF and CDF of the end-to-end  SNR, and developed  outage probability, average BER, and ergodic capacity  under the combined effect of cascaded oceanic turbulence and pointing errors in terms of Meijer's G and Fox's H functions.  We also provided the asymptotic expressions using Gamma functions for the outage probability and average BER to determine the diversity order of the considered system.  We validated our derived expressions using Monte-Carlo simulations and demonstrated the performance of UWOC under various channel conditions. The reported results may provide better performance assessment and design criteria of optical communication for higher underwater transmission links.
\section*{Appendix A}
To derive \eqref{eq:pdf_hc_hp} and \eqref{eq:cdf_hc_hp}, first, we need the PDF and CDF of $N$ cascaded channels $h_c=\prod_{i=1}^{N}h_{i}$.  We use the inverse Mellin transform to find the PDF of $h_c$. If $\mathbb{E}[X^n]$ denotes the $n$-th moment, where $\mathbb{E}[\cdot]$ denotes the expectation operator, then the inverse Mellin transform results the PDF of a random variable $X$ as
\begin{equation}
	f_X(x)=\frac{1}{2\pi ix}\int_{\zeta-i\infty}^{\zeta+i\infty}x^{-n}\mathbb{E}[X^n]dn
	\label{eq:Nth_order_with_Inverse_Mellin}
\end{equation}
where $\zeta-i\infty$ to $\zeta+i\infty$ denotes the line integral. It should be mentioned that Mellin transform has been used to analyze the product of $N$ random variable for different applications \cite{Kaddoum2018, Chapala2021,Chapala2021_LCOMM,bhardwaj2021performance}. The $n$-th order moment for $h_c$ is derived as	
\begin{eqnarray}
	&\mathbb{E}[h_c^n]=\prod_{i=1}^{N}\int_{0}^{\infty}h_i^n f_{h_i}(h_i)dh_i
	\label{eq:nth_order_moment_of_hi}
\end{eqnarray}
Substituting \eqref{eq:h_gg} in \eqref{eq:nth_order_moment_of_hi} and applying the identity $\int_{0}^{\infty}x^me^{-\beta x^n}dx=\frac{\Gamma\left(\frac{m+1}{n}\right)}{n\beta^{\frac{m+1}{n}}}$ \cite[pp. $347$, eq. $3.381.10$]{Zwillinger2014}, we get the $n$-th  moment of $h_c$ as:
\begin{equation}
	\mathbb{E}[h_c^n]=\prod_{i=1}^{N} \frac{\Gamma\left(\frac{n+d_i}{p_i}\right)}{a_i^{-n}\Gamma\left(\frac{d_i}{p_i}\right)}
	\label{eq:nth_order_moment_of_h_gen_g}
\end{equation}
Next, we use \eqref{eq:nth_order_moment_of_h_gen_g} in \eqref{eq:Nth_order_with_Inverse_Mellin} and apply the definition of Fox H-function to get the PDF of $h_c$ as 
\begin{eqnarray}
	&f_{h_c}(h_c)=\prod_{i=1}^{N} \frac{1}{h_c\Gamma\left(\frac{d_i}{p_i}\right)}\nonumber \\& H_{0,N}^{N,0} \left[\begin{matrix} - \\\left\{\left(\frac{d_{i_1}}{p_{i_1}},\frac{1}{p_{i_1}} \right)\right\}_{i_1=1}^{N}  \end{matrix} \bigg|\prod_{i_2=1}^{N} \left(\frac{h_c}{a_{i_2}}\right)\right]
	\label{eq:PDF_channel_Fox_H}
\end{eqnarray}

Next, we use the theory of product distribution  \cite{Papoulis2001} to get the PDF of the combined channel $h=h_ch_p$ as
\begin{eqnarray}
	f_{h}(h)=\int_{0}^{A_0}\frac{1}{h_p}f_{h_p}(h_p)f_{h_c}\left(\frac{h}{h_p} \right)dh_p
	\label{eq:combined_hc_hp}
\end{eqnarray}

Using \eqref{eq:pdf_hp} and \eqref{eq:PDF_channel_Fox_H} in \eqref{eq:combined_hc_hp} and applying the definition of Fox H-function, we get
\begin{eqnarray}
	&f_{h}(h)=\frac{\rho^2}{h}\prod_{i=1}^{N} \frac{1}{\Gamma\left(\frac{d_i}{p_i}\right)} H_{1,N+1}^{N+1,0} \nonumber \\&\left[\begin{matrix} (1+\rho^2,1) \\\left\{\left(\frac{d_{i_1}}{p_{i_1}},\frac{1}{p_{i_1}} \right)\right\}_{i_1=1}^{N},(\rho^2,1) \end{matrix} \bigg|\prod_{i_2=1}^{N} \left(\frac{h}{a_{i_2}A_0}\right)\right]
	\label{eq:pdf_hc_hp2} 
\end{eqnarray}

Finally, we use the transformation of random variable $\gamma= \gamma_0 h^2$ to get the PDF SNR in \eqref{eq:pdf_hc_hp}. To find the CDF of SNR under the combined channel, we use \eqref{eq:pdf_hc_hp} in $ F_{\gamma}(\gamma)=\int_0^{\gamma}f(\gamma)d\gamma$
and apply the definition of Fox H-function with inner integral $\int_{0}^{\gamma}\gamma^{-\frac{u}{2}-1} d\gamma=\frac{\gamma^{-\frac{u}{2}}}{-\frac{u}{2}}=\frac{2\gamma^{-\frac{u}{2}}\Gamma(-u)}{\Gamma(1-u)}$ to get the CDF of the SNR in \eqref{eq:cdf_hc_hp}, which concludes the proof of Theorem 1.
\bibliographystyle{ieeetran}
\typeout{} 
\bibliography{bib_file}
\end{document}